\documentclass[prmaterials,preprint]{revtex4-1}

\usepackage{graphicx}  

\usepackage{color}

\begin{document}

\title{Effective reduction of PdCoO$_2$ thin films via hydrogenation and sign tunable anomalous Hall effect}

\author{Gaurab Rimal}
\email{gr380@physics.rutgers.edu}
\affiliation{Department of Physics \& Astronomy, Rutgers, The State University of New Jersey, Piscataway, New Jersey 08854, USA }

\author{Yiting Liu}
\affiliation{Department of Physics \& Astronomy, Rutgers, The State University of New Jersey, Piscataway, New Jersey 08854, USA }
\affiliation{State Key Laboratory of Precision Spectroscopy, East China Normal University, Shanghai 200062, China }

\author{Caleb Schmidt}
\author{Hussein Hijazi}
\affiliation{Department of Physics \& Astronomy, Rutgers, The State University of New Jersey, Piscataway, New Jersey 08854, USA }

\author{Elizabeth Skoropata}
\author{Jason Lapano}
\affiliation{Materials Science and Technology Division, Oak Ridge National Laboratory,Oak Ridge, Tennessee 37831, USA }

\author{Debangshu Mukherjee}
\author{Raymond R. Unocic}
\author{Matthew F. Chisholm}
\affiliation{Center for Nanophase Materials Sciences, Oak Ridge National Laboratory, Oak Ridge, Tennessee 37831, USA }

\author{Yifei Sun}
\author{Haoming Yu}
\affiliation{School of Materials Engineering, Purdue University, West Lafayette, Indiana 47907, USA }

\author{Cheng-Jun Sun}
\author{Hua Zhou}
\affiliation{X-ray Science Division, Advanced Photon Source, Argonne National Laboratory, Lemont, Illinois 60439, USA}

\author{Matthew Brahlek}
\affiliation{Materials Science and Technology Division, Oak Ridge National Laboratory,Oak Ridge, Tennessee 37831, USA }

\author{Leonard C. Feldman} 
\affiliation{Department of Physics \& Astronomy, Rutgers, The State University of New Jersey, Piscataway, New Jersey 08854, USA }

\author{Shriram Ramanathan}
\affiliation{School of Materials Engineering, Purdue University, West Lafayette, Indiana 47907, USA }

\author{Seongshik Oh}
\email{ohsean@physics.rutgers.edu}
\affiliation{Department of Physics \& Astronomy, Rutgers, The State University of New Jersey, Piscataway, New Jersey 08854, USA }
\affiliation{Center for Quantum Materials Synthesis, Rutgers, The State University of New Jersey, Piscataway, New Jersey 08854, USA }



\begin{abstract}
	PdCoO$_2$, belonging to a family of triangular oxides called delafossite, is one of the most conducting oxides. Its in-plane conductivity is comparable to those of the best metals, and exhibits hydrodynamic electronic transport with extremely long mean free path at cryogenic temperatures. Nonetheless, it is nonmagnetic despite the presence of the cobalt ion. Here, we show that a mild hydrogenation process reduces PdCoO$_2$ thin films to an atomically-mixed alloy of PdCo with strong out-of-plane ferromagnetism and sign-tunable anomalous Hall effect. Considering that many other compounds remain little affected under a similar hydrogenation condition, this discovery may provide a route to creating novel spintronic heterostructures combining strong ferromagnetism, involving oxides and other functional materials. 
\end{abstract}

\maketitle



\begin{figure}[ht]
	\centering
	\includegraphics[width=0.75\textwidth]{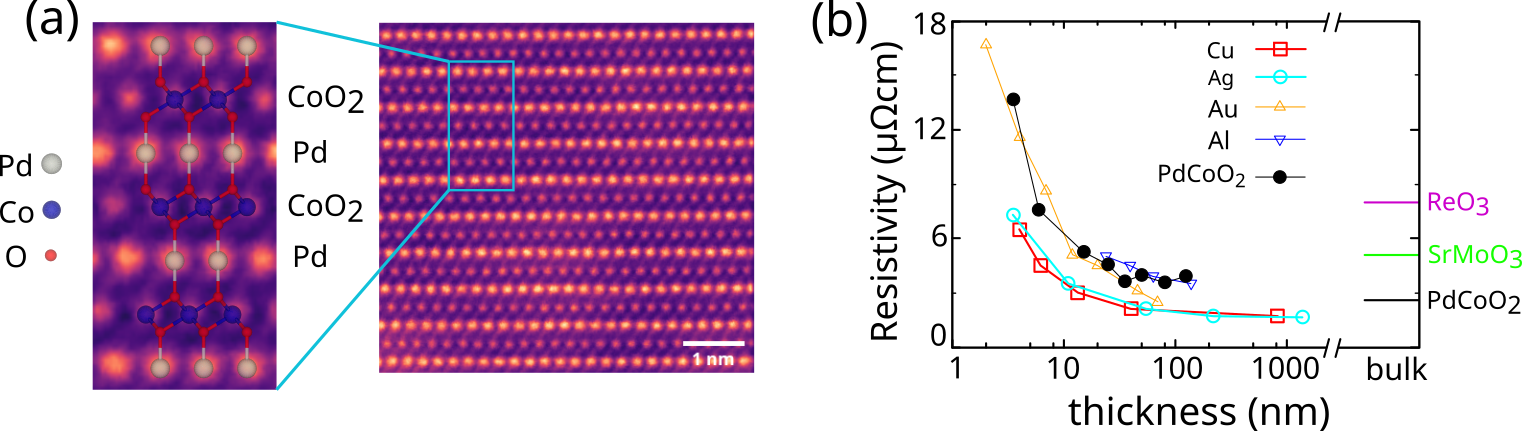}
	\caption{Structure and room-temperature resistivity of PdCoO$_2$ films. (a) Annular dark-field scanning transmission electron microscopy image for a PdCoO$_2$ thin film, with a zoomed region showing the atomic layering. Also overlaid is the unit cell aligned to proper atomic sites. (b) Comparison of room temperature resistivity for thin films of various metals (open markers), and PdCoO$_2$. The solid horizontal lines show the bulk value for the labeled oxides, taken from ref. \cite{Mackenzie2017,Nagai2005}. For metals, the resistivity values for Au were taken from ref. \cite{Kastle2004}, Cu and Ag from ref. \cite{Chawla2012}, and Al from ref. \cite{deVries1988}.} 
	\label{fig0}
\end{figure}

The delafossite family of oxides, named after the mineral CuFeO$_2$, with general molecular formula of ABO$_2$ has a three-fold layered crystal structure. The structure can be simply considered as alternating layers of A and BO$_2$ triangular lattice along the c-axis. The stacking sequence of the layers could result in either rhombohedral  ($R\bar{3}m$), as shown in Fig. \ref{fig0}(a), or hexagonal ($P6_3/mmc$) crystal structure. In general, the conduction properties are determined by the monovalent A site. When the A site is occupied by Cu or Ag, the system is usually insulating or semiconducting. On the other hand, Pt or Pd in the A site renders a metallic system. The trivalent B site, however, does not provide charge carriers, but acts to develop magnetism in some delafossites.

The metallic delafossites do not occur naturally. Although they were first synthesized in 1971 \cite{Shannon1971,Prewitt1971}, research activities were sparse in the following decades. More recently, however, as the quality of bulk crystals improved, the Pd/Pt based metallic system started to attract significant interest due to their unique transport properties. For example, in PdCoO$_2$, Pd is in an unusual 1+ oxidation state and provides one itinerant electron per site, giving an electron density of 1.45$\times$10$^{15}$ /cm$^2$ per Pd layer. The neighboring CoO$_2$ layer is, on the other hand, insulating. The alternation of conducting Pd layer and insulating CoO$_2$ layer results in highly anisotropic conductivity \cite{Takatsu2007,Daou2015,Yordanov2019}, and exotic transport properties such as hydrodynamic transport and ultra-low in-plane resistivity \cite{Moll2016}. Furthermore, as shown in Fig. \ref{fig0}(b), the room temperature resistivity of PdCoO$_2$ films is comparable to that of the best metals, which may be useful in technologies requiring highly conductive metallic oxides, such as interconnects.

Despite these many intriguing properties of PdCoO$_2$, it is non-magnetic because the 3d$^6$ electrons in Co$^{3+}$ ion completely fill the t$_{2g}$ band. At best, only weak signatures of spin-polarization have been observed on the surface of PdCoO$_2$, presumably due to surface ions with incomplete bonds \cite{Mazzola2017,Harada2020}. Other delafossites are also either nonmagnetic or weakly antiferromagnetic \cite{Mackenzie2017}. On the other hand, a related material, Pd/Co multilayer, can become a strong ferromagnet with perpendicular magnetic anisotropy (PMA) when the thickness of each layer is optimized. Despite the compositional similarities between PdCoO$_2$ and Pd/Co multilayers, these two materials have so far been investigated separately by two different communities with little connections between the two. Here, we show that with a mild hydrogenation process, PdCoO$_2$ can be easily reduced to PdCo alloys with strong PMA, and additional tunabilities can be incorporated. 

Hydrogenation has long been used as a way to modify the properties of oxides as well as semiconductors. In transition metal oxides, various hydrogenation schemes have been utilized to develop new functional properties by reducing the valence state of the transition metal \cite{Shi2014,Ramadoss2016,Li2019}. Nonetheless, most of these oxides require severe hydrogenation conditions or a catalyst such as Pd or Pt (for hydrogen spillover) in order to observe notable changes in their properties. In the present study, however, we find that even in a mild hydrogenation condition, annealing at 100 $\sim$ 200 $^\circ$C in a flow of diluted (10\%) hydrogen gas, PdCoO$_2$ films completely reduce to PdCo alloys with only trace amounts of oxygen and hydrogen. All the PdCoO$_2$ films used in this study were grown on Al$_2$O$_3$ (0001) substrates as described in Ref. \cite{Brahlek2019}, and further experimental details are outlined in supplemental material. All films were 9 nm thick, and hydrogenation conditions were anneal temperature (T$_A$) = 200 $^\circ$C, anneal time (t$_A$) = 30 min, unless indicated. 

\begin{figure}[ht]
	\centering
	\includegraphics[width=0.8\textwidth]{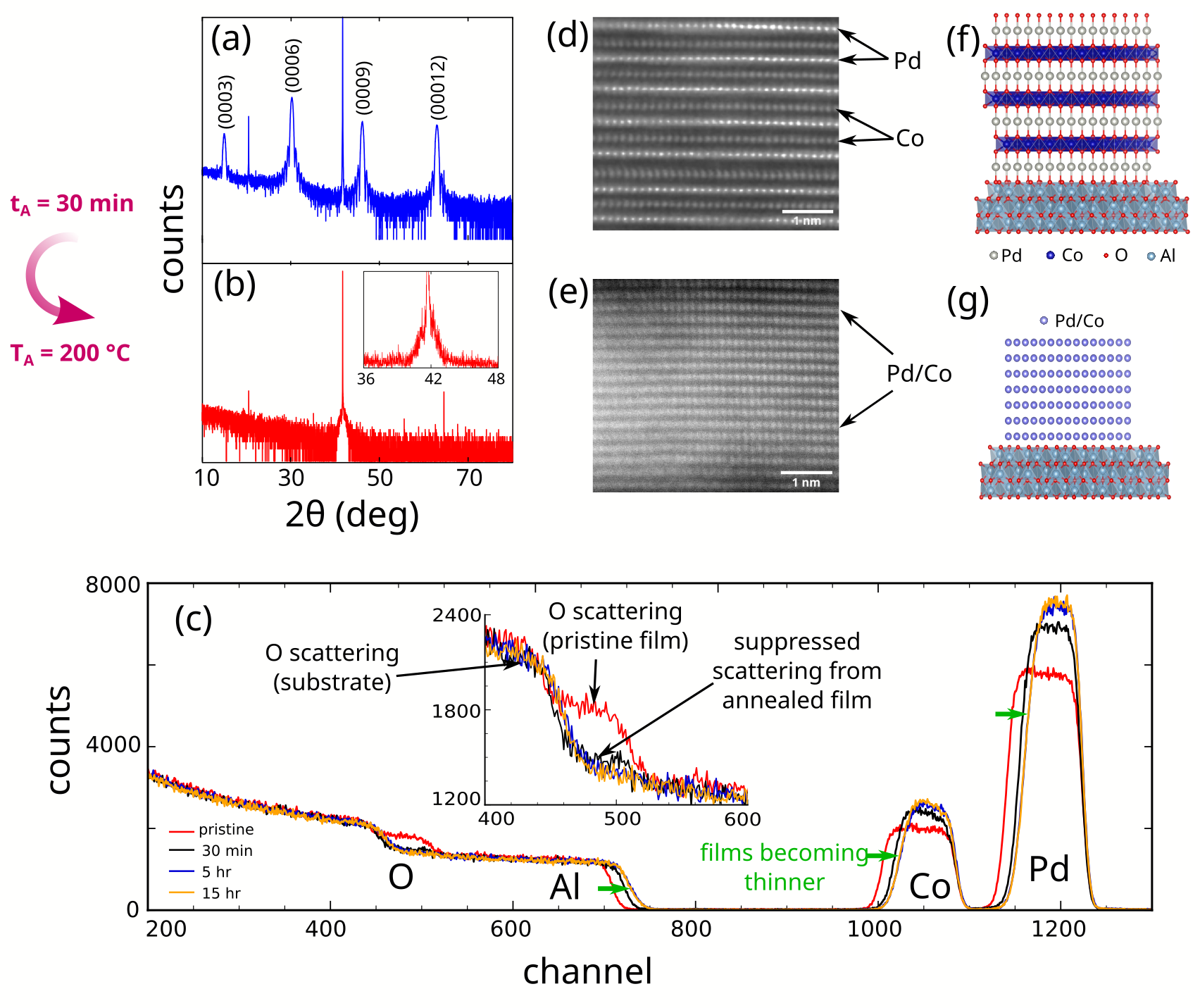}
	\caption{Structural comparison of pristine and hydrogenated PdCoO$_2$ films. XRD for 9 nm thick PdCoO$_2$ (a) before and (b) after hydrogenation. (c) RBS spectra for 100 nm thick films. The elements corresponding to features in spectra are marked. The inset shows oxygen region corresponding to backscattering contributions from the film and the substrate. The shift in edges of the backscattering features for various elements, as indicated by the arrows, is due to the reduction of film thickness. The approximate overlap of the curves for the 5 hr and 15 hr annealed curves indicates that the reduction process reached its limit of negligible oxygen content (see also supplemental material Fig. S1). (d) STEM image for a pristine PdCoO$_2$  film, showing atomic layering. (e) STEM for a 9 nm thick film hydrogenated at 200 $^\circ$C for 30 mins. The lack of contrast between atoms and layers is due to intermixing of Pd and Co. (f) Schematic structure of PdCoO$_2$ and (g) hydrogenated PdCoO$_2$. }
\label{fig1}
\end{figure}

In Fig. \ref{fig1}, we show the structural changes of the PdCoO$_2$ films after hydrogenation. X-ray diffraction (XRD) in Figs. \ref{fig1}(a) and \ref{fig1}(b) show that after hydrogenation, the sharp peaks corresponding to (000l) planes of the delafossite phase collapse to a single, broad peak overlapping with the Al$_2$O$_3$ (0006) peak: this single XRD peak lies between the diffraction angles that would correspond to Pd and Co (111) peaks \cite{Shaw2009}. The lack of satellite peaks also suggests that the multilayer has collapsed into an alloyed phase \cite{Weller1993}. Compared with the nearest Pd-Co layer spacing of 2.95 $\mathrm{\AA}$ in PdCoO$_2$, the out-of-plane lattice spacing after hydrogenation is 2.18 $\pm$ 0.02 $\mathrm{\AA}$. Thus, XRD clearly suggests that after hydrogenation the delafossite structure collapses into a new phase with reduced lattice constant and similar in-plane symmetry, albeit with some disorder.

In Fig. \ref{fig1}(c) we compare Rutherford backscattering (RBS) spectra for 100 nm thick films of PdCoO$_2$ after various anneal durations. In RBS spectra, backscattering due to different elements provides intensity contributions corresponding to their atomic mass. Since Pd and Co are found only in the film, they appear as a peak/stepped plateau in the spectrum. However, Al and O, which are found in the much thicker substrate, show a continuous feature at lower energies (channels). The inset shows an additional feature in the oxygen region for the pristine film, which starts at a slightly higher energy than the oxygen signal from the substrate. This extra intensity arises from oxygen in the film, which decreases and eventually drops below the detection limit after extended annealing, as discussed more in depth in supplemental material (Fig. S1). Furthermore, the loss of oxygen causes decrease in the film thickness, which results in the decrease of peak widths of Pd and Co. Nonetheless, the area under the features of Pd and Co stay unchanged, implying that the content of Pd and Co remains the same. As outlined in Table S1, the hydrogen content also reduces as anneal time is increased, which suggests that the main role of hydrogen is to remove the oxygen away from the sample rather than bond to either Pd or Co ions.

The collapse of the delafossite structure and reduction of the films is evident from the scanning transmission electron microscopy (STEM) images shown in Figs. \ref{fig1}(d) and \ref{fig1}(e). The in-plane lattice decreases from 2.83 $\mathrm{\AA}$ to 2.5 $\pm$ 0.2 $\mathrm{\AA}$ and the inter-planar distance decreases from 2.95 $\mathrm{\AA}$ to 2.3 $\pm$ 0.1 $\mathrm{\AA}$. Furthermore, while the delafossite phase shows a clear contrast between the Pd and Co layers, this information is lost in the hydrogenated film, suggesting significant intermixing between Pd and Co. This collapse is schematically shown in Figs. \ref{fig1}(f) and \ref{fig1}(g). A recent study of the catalytic activities of PdCoO$_2$ bulk single crystals has also found significant changes to the surface due to hydrogen evolution \cite{Li2019Felser}, and the role of hydrogen in our films is possibly similar. To determine the nature of the collapse of PdCoO$_2$ due to hydrogenation, we employed x-ray absorption (XAS), involving both near edge spectroscopy (XANES) and extended x-ray absorption fine structure spectroscopy (EXAFS) to probe the variation of local environment around Pd ions.

\begin{figure}[ht]
	\centering
	\includegraphics[width=0.65\textwidth]{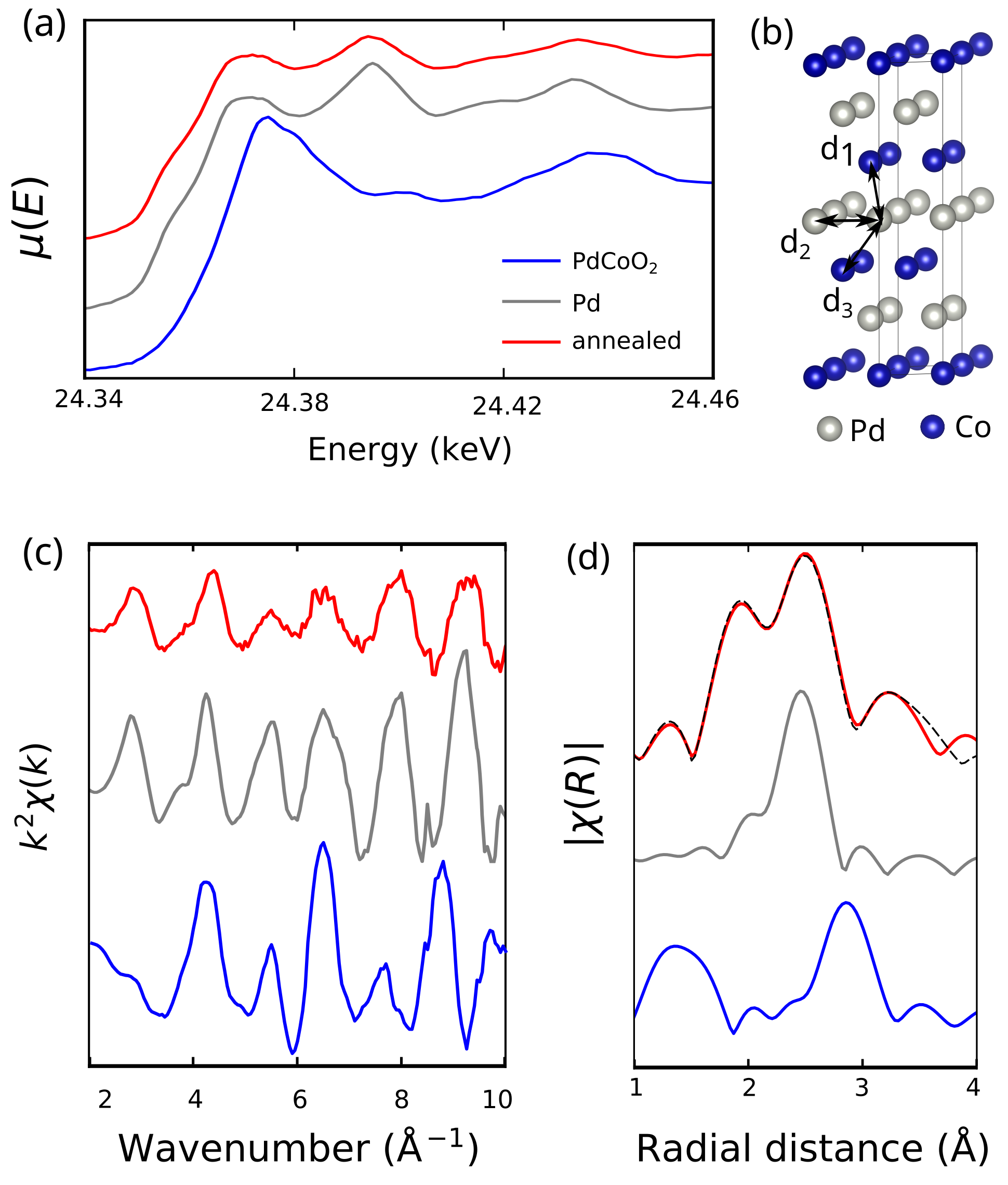}
	\caption{XAS comparisons between pristine PdCoO$_2$, Pd and hydrogenated PdCoO$_2$ with T$_A$ = 400 $^\circ$C and t$_A$ = 30 min. (a) Pd K-edge XAS spectra in the near-edge region. The main edge for hydrogenated sample is red-shifted from the pristine PdCoO$_2$, and overlaps with Pd metal. (b) PdCo model for fitting EXAFS employing $R\bar{3}m$ structure. d$_1$, d$_2$ and d$_3$ are the nearest (Co), second-nearest (Pd) and third-nearest (Co) atomic distances to the absorbing Pd site. (c) k$^2$ weighted EXAFS showing the fine structure of interference due to nearest neighbor scattering. (d) Magnitude of radial distance dependence of EXAFS. The best fit for hydrogenated PdCoO$_2$ at 400 $^\circ$C using model shown in (b) is also shown as the dashed line.  }
	\label{fig2}
\end{figure}

In Fig. \ref{fig2}, we show the results of Pd K-edge XAS on nominally 9 nm thick PdCoO$_2$, before and after hydrogenation. As evident in Fig. \ref{fig2}(a), hydrogenation shifts the absorption edge to lower energies, which overlaps with the edge of Pd metal, thus confirming the reduced nature of the films. Furthermore, the XANES spectra for hydrogenated films annealed at various temperatures all align well with that for pure Pd metal, suggesting that the films are fully reduced to the pure elemental state (supplemental Fig. S2). However, as shown in Fig. \ref{fig2}(c,d), the hydrogenated films show different behavior in the local environment. The wavenumber and radial distance dependence of EXAFS parameter around the Pd ion are all different among pristine (PdCoO$_2$), hydrogenated and pure Pd. Since EXAFS is most sensitive  to the nearest neighbors, we determine the local environment around the absorbing Pd atoms in hydrogenated PdCoO$_2$ by fitting the spectra using an $R\bar{3}m$ PdCo structure with a 3-shell model shown in Fig. \ref{fig2}(b), in which Pd is surrounded by 6 nearest Co atoms, 6 second-nearest Pd and 6 third-nearest Co atoms. As shown by the fitted line in Fig. \ref{fig2}(d), the sample annealed at the highest temperature (400 $^\circ$C) is well described by this structure. However, the nearest Co and the second-nearest Pd are found to have similar distances (2.66 $\mathrm{\AA}$ and 2.72 $\mathrm{\AA}$, respectively), which means that there are likely 12 nearest Pd and Co neighbors surrounding a given Pd atom, at an average distance of 2.68 $\mathrm{\AA}$ and 6 next-nearest Co at 3.95 $\mathrm{\AA}$. In contrast, the films annealed at 100 $^\circ$C and 200 $^\circ$C cannot be fit by the model shown in Fig. \ref{fig2}(b). In fact, we were not able to find any reasonable model that describes these low temperature-annealed films, which could be due to residual oxygen and hydrogen, or incomplete mixing between Pd and Co, or both.  

The EXAFS results point to the case in which gradual reduction of PdCoO$_2$ helps retain the layering of Pd and Co. However, unlike other PdCo alloys and multilayers \cite{Weller1993,Kim1997,Harumoto2019,Takahashi1993}, in which the structure is generally fcc-derived, there may be microstructural differences in these reduced films. Although more experiments are needed to elucidate the exact structure of the reduced films, it is possible that the starting $R\bar{3}m$ structure of PdCoO$_2$ provides a route to obtaining PdCo alloys with different properties compared to films obtained by traditional deposition methods.

\begin{figure}[ht]
	\centering
	\includegraphics[width=0.95\textwidth]{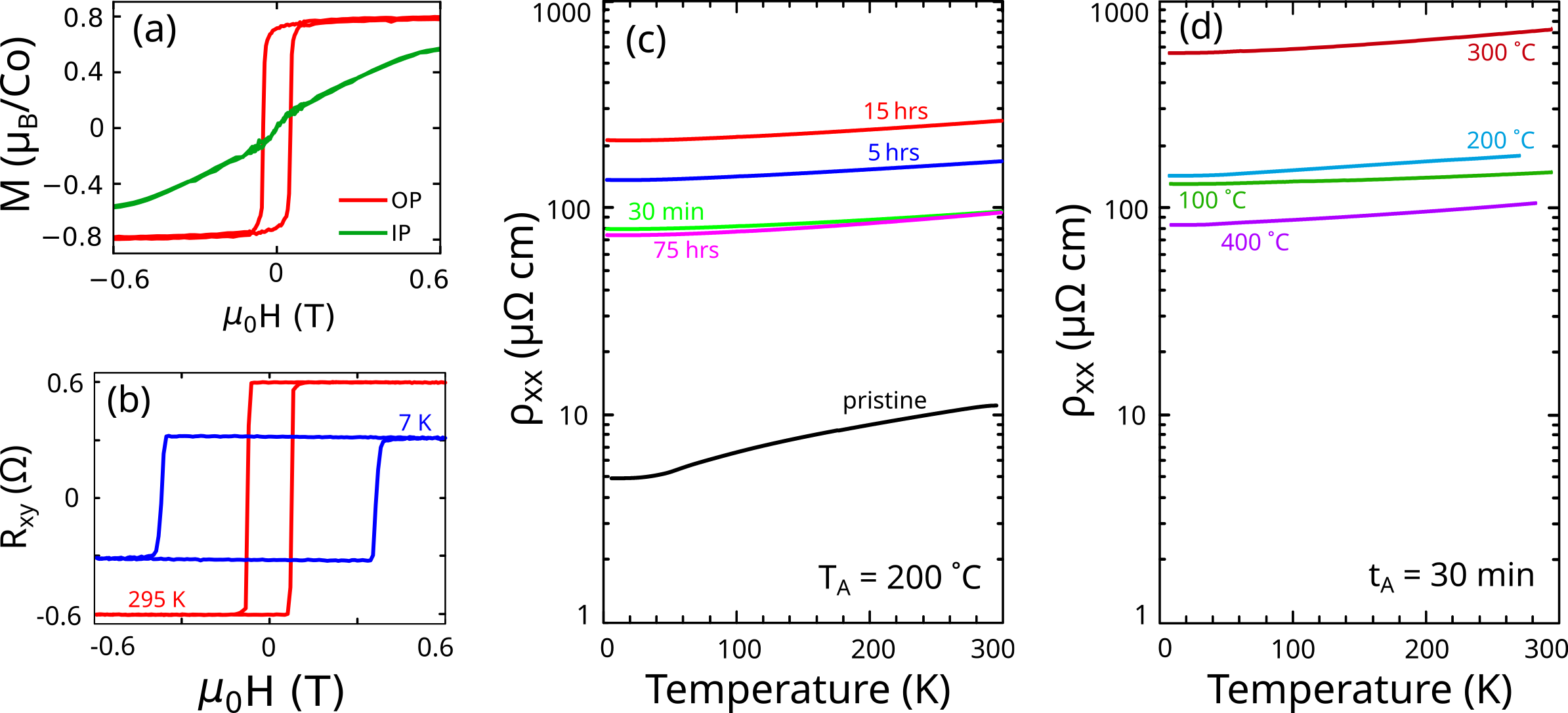}
	\caption{Magnetic and transport properties of 9 nm thick hydrogenated PdCoO$_2$ films. (a) Room temperature magnetic hysteresis comparing the in-plane (IP) and out-of-plane (OP) directions. (b) AHE curves at room temperature and low temperature showing the sharp transition corresponding to strong OP anisotropy. (c) Temperature dependence of resistivity for various anneal durations (t$_A$) at T$_A$ = 200 $^\circ$C. (d) Temperature dependence of resistivity for various anneal temperatures (T$_A$) with t$_A$ = 30 min.}
	\label{fig3}
\end{figure}

The reduced PdCoO$_2$ films also develop strong ferromagnetism with PMA, as indicated by the magnetization loop in Fig. \ref{fig3}(a) and anomalous Hall effect (AHE) in Fig. \ref{fig3}(b). The estimated Curie temperature is $\sim$650 K, as shown in the temperature dependent magnetization data in supplemental Fig. S3. According to previous studies \cite{Wu1992,Weller1994,Hong2005}, it is generally believed that the multilayer structure of the Pd/Co system is critical for achieving strong PMA. However, the very observation of strong PMA in this atomically intermixed PdCo system suggests that it is less the layering but more close proximity between Pd and Co ions that is critical for strong PMA. 

The longitudinal resistivity ($\rho_{xx}$) in Figs. \ref{fig3}(c) and \ref{fig3}(d) also show systematic changes with the level of hydrogenation: it first increases with hydrogenation, then reaches a maximum, and finally falls to a smaller value. This general trend can be understood as follows. As the oxygen is removed and the delafossite structure collapses, the characteristic low resistivity values of PdCoO$_2$ films increase toward modest values of the Pd-Co alloys. However, after extensive hydrogenation (with t$_A$ = 75 hrs, or T$_A$ = 400 $^\circ$C), as the traces of oxygen and hydrogen gradually disappear, accompanied by maximal intermixing between Pd and Co, electron scattering will reduce, resulting in the lower resistivity, albeit much larger than that of the original PdCoO$_2$ system. 

\begin{figure}[ht]
	\centering
	\includegraphics[width=0.6\textwidth]{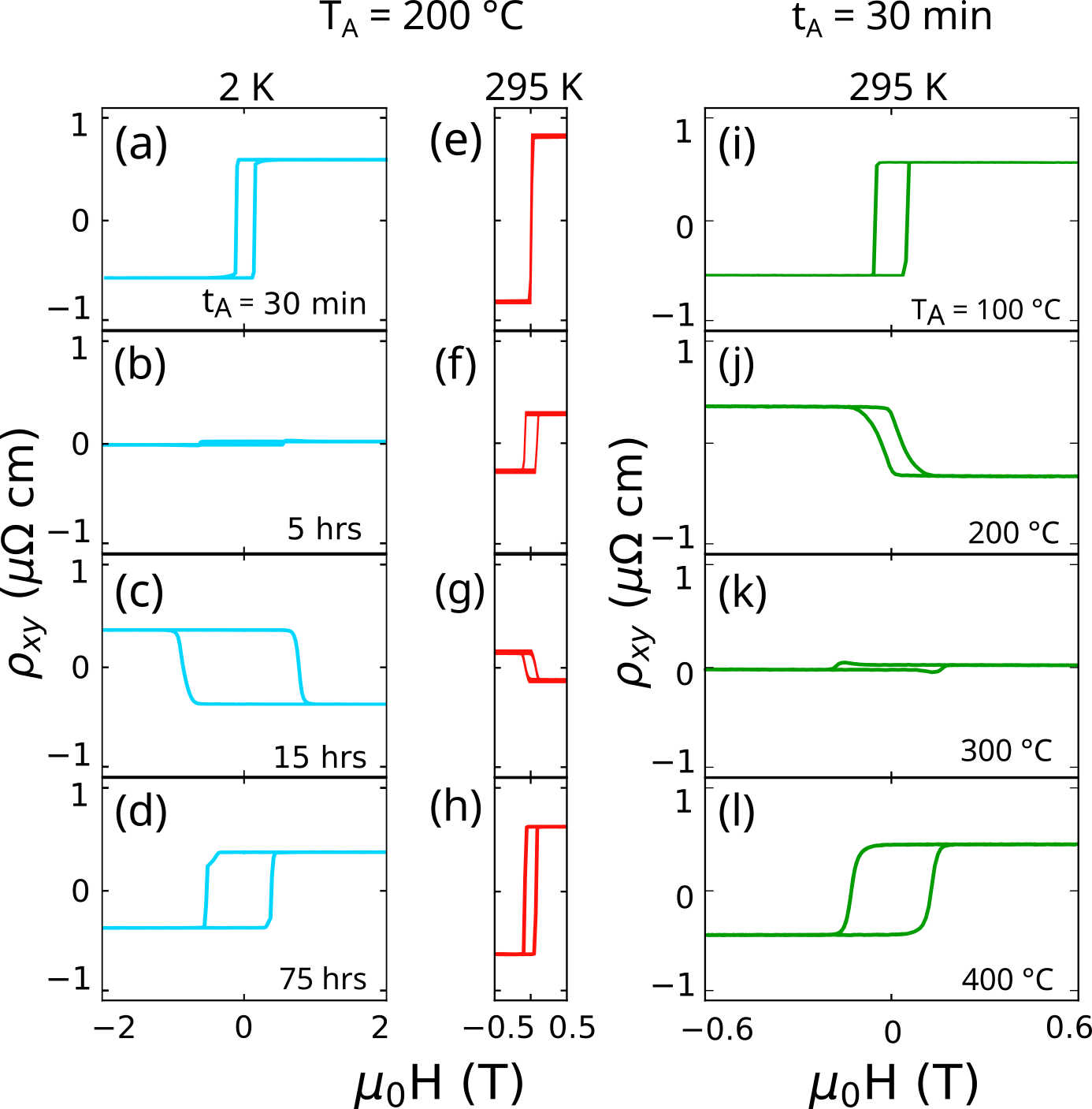}
	\caption{AHE for hydrogenated films with various anneal parameters. (a-d) AHE for 9 nm thick films at 2 K for T$_A$ = 200 $^\circ$C for various anneal durations. (e-h) AHE for 9 nm thick films at 295 K for T$_A$ = 200 $^\circ$C for various anneal durations. (i-l) AHE for films annealed at various temperatures for t$_A$ = 30 min. }
	\label{fig4}
\end{figure}

There are even more dramatic changes in the transverse (Hall) resistivity, $\rho_{xy}$, with the level of hydrogenation, as shown in Fig. \ref{fig4}: not only the magnitude but also the sign of the AHE changes with hydrogenation time and temperature. In Figs. \ref{fig4}(a-d), the magnitude of $\rho_{xy}$ initially decreases, but after annealing for 15 hrs, the sign reverses. With further anneal, the sign reverts back to the original. Similar behavior is also observed with anneal temperature, shown in Figs. \ref{fig4}(i-l). It is also notable that despite all these changes in AHE, carrier type and density, as estimated by the high field slopes, remain almost unchanged. Such switching behaviors were observed on eight different samples, indicating the reproducibility of this process.  In case of Pd/Co multilayers, different AHE sign requires different layer thicknesses \cite{Kim1993,Aoki1996,Rosenblatt2010,Keskin2013,Winer2015}.  What is unique to the reduced PdCoO$_2$ system, is that its AHE sign can be controlled by the level of hydrogenation. With hydrogenation, the layered PdCoO$_2$ structure gradually evolves toward atomically-intermixed PdCo alloys, which suggests that the AHE sign is a sensitive function of the structure including the level of intermixing between Pd and Co ions in the PdCo system.

This work shows that the highly conducting, non-magnetic PdCoO$_2$ films can be controllably reduced to an atomically-mixed PdCo system with strong out-of-plane ferromagnetism and sign-tunable AHE using a mild hydrogenation process. The fact that PdCoO$_2$ shares the 3-fold in-plane symmetry with many quantum materials such as graphene, 2D chalcogenides and other topological materials, further opens the possibility of forming various hybrid structures for broad applications. Moreover, considering that many oxides and chalcogenides remain intact during such a mild hydrogenation condition, a selective hydrogenation process could be envisioned to create unique magneto-electric heterostructures that cannot be created by other methods.



\section*{Acknowledgements} 
This work is mainly supported by National Science Foundation (NSF) Grant No. DMR2004125 and Army Research Office (ARO) Grant No. W911NF-20-1-0108. Y.L. acknowledges the state scholarship provided by the China scholarship council. E.S., J.L., and M.B. acknowledge the support by the U.S. Department of Energy (DOE), Office of Science, Basic Energy Sciences, Materials Sciences and Engineering Division (magnetic measurements and structural characterization). S.R. acknowledges National Science Foundation Award 1904081. STEM imaging was conducted at the Center for Nanophase Materials Sciences, which is a DOE office of science user facility. This research used resources of the Advanced Photon Source, a U.S. Department of Energy (DOE) Office of Science User Facility, operated for the DOE Office of Science by Argonne National Laboratory under Contract No. DE-AC02-06CH11357. Extraordinary facility operations were supported in part by the DOE Office of Science through the National Virtual Biotechnology Laboratory, a consortium of DOE national laboratories focused on the response to COVID-19, with funding provided by the Coronavirus CARES Act.


\begin{thebibliography}{31}%
\makeatletter
\providecommand \@ifxundefined [1]{%
 \@ifx{#1\undefined}
}%
\providecommand \@ifnum [1]{%
 \ifnum #1\expandafter \@firstoftwo
 \else \expandafter \@secondoftwo
 \fi
}%
\providecommand \@ifx [1]{%
 \ifx #1\expandafter \@firstoftwo
 \else \expandafter \@secondoftwo
 \fi
}%
\providecommand \natexlab [1]{#1}%
\providecommand \enquote  [1]{``#1''}%
\providecommand \bibnamefont  [1]{#1}%
\providecommand \bibfnamefont [1]{#1}%
\providecommand \citenamefont [1]{#1}%
\providecommand \href@noop [0]{\@secondoftwo}%
\providecommand \href [0]{\begingroup \@sanitize@url \@href}%
\providecommand \@href[1]{\@@startlink{#1}\@@href}%
\providecommand \@@href[1]{\endgroup#1\@@endlink}%
\providecommand \@sanitize@url [0]{\catcode `\\12\catcode `\$12\catcode
  `\&12\catcode `\#12\catcode `\^12\catcode `\_12\catcode `\%12\relax}%
\providecommand \@@startlink[1]{}%
\providecommand \@@endlink[0]{}%
\providecommand \url  [0]{\begingroup\@sanitize@url \@url }%
\providecommand \@url [1]{\endgroup\@href {#1}{\urlprefix }}%
\providecommand \urlprefix  [0]{URL }%
\providecommand \Eprint [0]{\href }%
\providecommand \doibase [0]{http://dx.doi.org/}%
\providecommand \selectlanguage [0]{\@gobble}%
\providecommand \bibinfo  [0]{\@secondoftwo}%
\providecommand \bibfield  [0]{\@secondoftwo}%
\providecommand \translation [1]{[#1]}%
\providecommand \BibitemOpen [0]{}%
\providecommand \bibitemStop [0]{}%
\providecommand \bibitemNoStop [0]{.\EOS\space}%
\providecommand \EOS [0]{\spacefactor3000\relax}%
\providecommand \BibitemShut  [1]{\csname bibitem#1\endcsname}%
\let\auto@bib@innerbib\@empty
\bibitem [{\citenamefont {Shannon}\ \emph {et~al.}(1971)\citenamefont
  {Shannon}, \citenamefont {Rogers},\ and\ \citenamefont
  {Prewitt}}]{Shannon1971}%
  \BibitemOpen
  \bibfield  {author} {\bibinfo {author} {\bibfnamefont {R.~D.}\ \bibnamefont
  {Shannon}}, \bibinfo {author} {\bibfnamefont {D.~B.}\ \bibnamefont {Rogers}},
  \ and\ \bibinfo {author} {\bibfnamefont {C.~T.}\ \bibnamefont {Prewitt}},\
  }\href {\doibase 10.1021/ic50098a011} {\bibfield  {journal} {\bibinfo
  {journal} {Inorganic Chemistry}\ }\textbf {\bibinfo {volume} {10}},\ \bibinfo
  {pages} {713} (\bibinfo {year} {1971})}\BibitemShut {NoStop}%
\bibitem [{\citenamefont {Prewitt}\ \emph {et~al.}(1971)\citenamefont
  {Prewitt}, \citenamefont {Shannon},\ and\ \citenamefont
  {Rogers}}]{Prewitt1971}%
  \BibitemOpen
  \bibfield  {author} {\bibinfo {author} {\bibfnamefont {C.~T.}\ \bibnamefont
  {Prewitt}}, \bibinfo {author} {\bibfnamefont {R.~D.}\ \bibnamefont
  {Shannon}}, \ and\ \bibinfo {author} {\bibfnamefont {D.~B.}\ \bibnamefont
  {Rogers}},\ }\href {\doibase 10.1021/ic50098a012} {\bibfield  {journal}
  {\bibinfo  {journal} {Inorganic Chemistry}\ }\textbf {\bibinfo {volume}
  {10}},\ \bibinfo {pages} {719} (\bibinfo {year} {1971})}\BibitemShut
  {NoStop}%
\bibitem [{\citenamefont {Takatsu}\ \emph {et~al.}(2007)\citenamefont
  {Takatsu}, \citenamefont {Yonezawa}, \citenamefont {Mouri}, \citenamefont
  {Nakatsuji}, \citenamefont {Tanaka},\ and\ \citenamefont
  {Maeno}}]{Takatsu2007}%
  \BibitemOpen
  \bibfield  {author} {\bibinfo {author} {\bibfnamefont {H.}~\bibnamefont
  {Takatsu}}, \bibinfo {author} {\bibfnamefont {S.}~\bibnamefont {Yonezawa}},
  \bibinfo {author} {\bibfnamefont {S.}~\bibnamefont {Mouri}}, \bibinfo
  {author} {\bibfnamefont {S.}~\bibnamefont {Nakatsuji}}, \bibinfo {author}
  {\bibfnamefont {K.}~\bibnamefont {Tanaka}}, \ and\ \bibinfo {author}
  {\bibfnamefont {Y.}~\bibnamefont {Maeno}},\ }\href {\doibase
  10.1143/JPSJ.76.104701} {\bibfield  {journal} {\bibinfo  {journal} {Journal
  of the Physical Society of Japan}\ }\textbf {\bibinfo {volume} {76}},\
  \bibinfo {pages} {104701} (\bibinfo {year} {2007})}\BibitemShut {NoStop}%
\bibitem [{\citenamefont {Daou}\ \emph {et~al.}(2015)\citenamefont {Daou},
  \citenamefont {Fr{\'{e}}sard}, \citenamefont {H{\'{e}}bert},\ and\
  \citenamefont {Maignan}}]{Daou2015}%
  \BibitemOpen
  \bibfield  {author} {\bibinfo {author} {\bibfnamefont {R.}~\bibnamefont
  {Daou}}, \bibinfo {author} {\bibfnamefont {R.}~\bibnamefont {Fr{\'{e}}sard}},
  \bibinfo {author} {\bibfnamefont {S.}~\bibnamefont {H{\'{e}}bert}}, \ and\
  \bibinfo {author} {\bibfnamefont {A.}~\bibnamefont {Maignan}},\ }\href
  {\doibase 10.1103/PhysRevB.91.041113} {\bibfield  {journal} {\bibinfo
  {journal} {Physical Review B}\ }\textbf {\bibinfo {volume} {91}},\ \bibinfo
  {pages} {041113(R)} (\bibinfo {year} {2015})}\BibitemShut {NoStop}%
\bibitem [{\citenamefont {Yordanov}\ \emph {et~al.}(2019)\citenamefont
  {Yordanov}, \citenamefont {Sigle}, \citenamefont {Kaya}, \citenamefont
  {Gruner}, \citenamefont {Pentcheva}, \citenamefont {Keimer},\ and\
  \citenamefont {Habermeier}}]{Yordanov2019}%
  \BibitemOpen
  \bibfield  {author} {\bibinfo {author} {\bibfnamefont {P.}~\bibnamefont
  {Yordanov}}, \bibinfo {author} {\bibfnamefont {W.}~\bibnamefont {Sigle}},
  \bibinfo {author} {\bibfnamefont {P.}~\bibnamefont {Kaya}}, \bibinfo {author}
  {\bibfnamefont {M.~E.}\ \bibnamefont {Gruner}}, \bibinfo {author}
  {\bibfnamefont {R.}~\bibnamefont {Pentcheva}}, \bibinfo {author}
  {\bibfnamefont {B.}~\bibnamefont {Keimer}}, \ and\ \bibinfo {author}
  {\bibfnamefont {H.-U.}\ \bibnamefont {Habermeier}},\ }\href {\doibase
  10.1103/physrevmaterials.3.085403} {\bibfield  {journal} {\bibinfo  {journal}
  {Physical Review Materials}\ }\textbf {\bibinfo {volume} {3}},\ \bibinfo
  {pages} {85403} (\bibinfo {year} {2019})}\BibitemShut {NoStop}%
\bibitem [{\citenamefont {Moll}\ \emph {et~al.}(2016)\citenamefont {Moll},
  \citenamefont {Kushwaha}, \citenamefont {Nandi}, \citenamefont {Schmidt},\
  and\ \citenamefont {Mackenzie}}]{Moll2016}%
  \BibitemOpen
  \bibfield  {author} {\bibinfo {author} {\bibfnamefont {P.~J.}\ \bibnamefont
  {Moll}}, \bibinfo {author} {\bibfnamefont {P.}~\bibnamefont {Kushwaha}},
  \bibinfo {author} {\bibfnamefont {N.}~\bibnamefont {Nandi}}, \bibinfo
  {author} {\bibfnamefont {B.}~\bibnamefont {Schmidt}}, \ and\ \bibinfo
  {author} {\bibfnamefont {A.~P.}\ \bibnamefont {Mackenzie}},\ }\href {\doibase
  10.1126/science.aac8385} {\bibfield  {journal} {\bibinfo  {journal}
  {Science}\ }\textbf {\bibinfo {volume} {351}},\ \bibinfo {pages} {1061}
  (\bibinfo {year} {2016})}\BibitemShut {NoStop}%
\bibitem [{\citenamefont {Mazzola}\ \emph {et~al.}(2017)\citenamefont
  {Mazzola}, \citenamefont {Sunko}, \citenamefont {Khim}, \citenamefont
  {Rosner}, \citenamefont {Kushwaha}, \citenamefont {Clark}, \citenamefont
  {Bawden}, \citenamefont {Markovi{\'{c}}}, \citenamefont {Kim}, \citenamefont
  {Hoesch}, \citenamefont {Mackenzie},\ and\ \citenamefont
  {King}}]{Mazzola2017}%
  \BibitemOpen
  \bibfield  {author} {\bibinfo {author} {\bibfnamefont {F.}~\bibnamefont
  {Mazzola}}, \bibinfo {author} {\bibfnamefont {V.}~\bibnamefont {Sunko}},
  \bibinfo {author} {\bibfnamefont {S.}~\bibnamefont {Khim}}, \bibinfo {author}
  {\bibfnamefont {H.}~\bibnamefont {Rosner}}, \bibinfo {author} {\bibfnamefont
  {P.}~\bibnamefont {Kushwaha}}, \bibinfo {author} {\bibfnamefont {O.~J.}\
  \bibnamefont {Clark}}, \bibinfo {author} {\bibfnamefont {L.}~\bibnamefont
  {Bawden}}, \bibinfo {author} {\bibfnamefont {I.}~\bibnamefont
  {Markovi{\'{c}}}}, \bibinfo {author} {\bibfnamefont {T.~K.}\ \bibnamefont
  {Kim}}, \bibinfo {author} {\bibfnamefont {M.}~\bibnamefont {Hoesch}},
  \bibinfo {author} {\bibfnamefont {A.~P.}\ \bibnamefont {Mackenzie}}, \ and\
  \bibinfo {author} {\bibfnamefont {P.~D.~C.}\ \bibnamefont {King}},\ }\href
  {\doibase 10.1073/pnas.1811873115} {\bibfield  {journal} {\bibinfo  {journal}
  {Proceedings of National Academy of Sciences}\ }\textbf {\bibinfo {volume}
  {115}},\ \bibinfo {pages} {12956} (\bibinfo {year} {2017})}\BibitemShut
  {NoStop}%
\bibitem [{\citenamefont {Harada}\ \emph {et~al.}(2020)\citenamefont {Harada},
  \citenamefont {Sugawara}, \citenamefont {Fujiwara}, \citenamefont {Kitamura},
  \citenamefont {Ito}, \citenamefont {Nojima}, \citenamefont {Horiba},
  \citenamefont {Kumigashira}, \citenamefont {Takahashi}, \citenamefont
  {Sato},\ and\ \citenamefont {Tsukazaki}}]{Harada2020}%
  \BibitemOpen
  \bibfield  {author} {\bibinfo {author} {\bibfnamefont {T.}~\bibnamefont
  {Harada}}, \bibinfo {author} {\bibfnamefont {K.}~\bibnamefont {Sugawara}},
  \bibinfo {author} {\bibfnamefont {K.}~\bibnamefont {Fujiwara}}, \bibinfo
  {author} {\bibfnamefont {M.}~\bibnamefont {Kitamura}}, \bibinfo {author}
  {\bibfnamefont {S.}~\bibnamefont {Ito}}, \bibinfo {author} {\bibfnamefont
  {T.}~\bibnamefont {Nojima}}, \bibinfo {author} {\bibfnamefont
  {K.}~\bibnamefont {Horiba}}, \bibinfo {author} {\bibfnamefont
  {H.}~\bibnamefont {Kumigashira}}, \bibinfo {author} {\bibfnamefont
  {T.}~\bibnamefont {Takahashi}}, \bibinfo {author} {\bibfnamefont
  {T.}~\bibnamefont {Sato}}, \ and\ \bibinfo {author} {\bibfnamefont
  {A.}~\bibnamefont {Tsukazaki}},\ }\href {\doibase
  10.1103/physrevresearch.2.013282} {\bibfield  {journal} {\bibinfo  {journal}
  {Physical Review Research}\ }\textbf {\bibinfo {volume} {2}},\ \bibinfo
  {pages} {013282} (\bibinfo {year} {2020})}\BibitemShut {NoStop}%
\bibitem [{\citenamefont {Mackenzie}(2017)}]{Mackenzie2017}%
  \BibitemOpen
  \bibfield  {author} {\bibinfo {author} {\bibfnamefont {A.~P.}\ \bibnamefont
  {Mackenzie}},\ }\href {\doibase 10.1088/1361-6633/aa50e5} {\bibfield
  {journal} {\bibinfo  {journal} {Reports on Progress in Physics}\ }\textbf
  {\bibinfo {volume} {80}},\ \bibinfo {pages} {032501} (\bibinfo {year}
  {2017})}\BibitemShut {NoStop}%
\bibitem [{\citenamefont {Shi}\ \emph {et~al.}(2014)\citenamefont {Shi},
  \citenamefont {Zhou},\ and\ \citenamefont {Ramanathan}}]{Shi2014}%
  \BibitemOpen
  \bibfield  {author} {\bibinfo {author} {\bibfnamefont {J.}~\bibnamefont
  {Shi}}, \bibinfo {author} {\bibfnamefont {Y.}~\bibnamefont {Zhou}}, \ and\
  \bibinfo {author} {\bibfnamefont {S.}~\bibnamefont {Ramanathan}},\ }\href
  {\doibase 10.1038/ncomms5860} {\bibfield  {journal} {\bibinfo  {journal}
  {Nature Communications}\ }\textbf {\bibinfo {volume} {5}},\ \bibinfo {pages}
  {4860} (\bibinfo {year} {2014})}\BibitemShut {NoStop}%
\bibitem [{\citenamefont {Ramadoss}\ \emph {et~al.}(2016)\citenamefont
  {Ramadoss}, \citenamefont {Mandal}, \citenamefont {Dai}, \citenamefont {Wan},
  \citenamefont {Zhou}, \citenamefont {Rokhinson}, \citenamefont {Chen},
  \citenamefont {Hu},\ and\ \citenamefont {Ramanathan}}]{Ramadoss2016}%
  \BibitemOpen
  \bibfield  {author} {\bibinfo {author} {\bibfnamefont {K.}~\bibnamefont
  {Ramadoss}}, \bibinfo {author} {\bibfnamefont {N.}~\bibnamefont {Mandal}},
  \bibinfo {author} {\bibfnamefont {X.}~\bibnamefont {Dai}}, \bibinfo {author}
  {\bibfnamefont {Z.}~\bibnamefont {Wan}}, \bibinfo {author} {\bibfnamefont
  {Y.}~\bibnamefont {Zhou}}, \bibinfo {author} {\bibfnamefont {L.}~\bibnamefont
  {Rokhinson}}, \bibinfo {author} {\bibfnamefont {Y.~P.}\ \bibnamefont {Chen}},
  \bibinfo {author} {\bibfnamefont {J.}~\bibnamefont {Hu}}, \ and\ \bibinfo
  {author} {\bibfnamefont {S.}~\bibnamefont {Ramanathan}},\ }\href {\doibase
  10.1103/PhysRevB.94.235124} {\bibfield  {journal} {\bibinfo  {journal}
  {Physical Review B}\ }\textbf {\bibinfo {volume} {94}},\ \bibinfo {pages}
  {235124} (\bibinfo {year} {2016})}\BibitemShut {NoStop}%
\bibitem [{\citenamefont {Li}\ \emph {et~al.}(2019{\natexlab{a}})\citenamefont
  {Li}, \citenamefont {Lee}, \citenamefont {Wang}, \citenamefont {Osada},
  \citenamefont {Crossley}, \citenamefont {Lee}, \citenamefont {Cui},
  \citenamefont {Hikita},\ and\ \citenamefont {Hwang}}]{Li2019}%
  \BibitemOpen
  \bibfield  {author} {\bibinfo {author} {\bibfnamefont {D.}~\bibnamefont
  {Li}}, \bibinfo {author} {\bibfnamefont {K.}~\bibnamefont {Lee}}, \bibinfo
  {author} {\bibfnamefont {B.~Y.}\ \bibnamefont {Wang}}, \bibinfo {author}
  {\bibfnamefont {M.}~\bibnamefont {Osada}}, \bibinfo {author} {\bibfnamefont
  {S.}~\bibnamefont {Crossley}}, \bibinfo {author} {\bibfnamefont {H.~R.}\
  \bibnamefont {Lee}}, \bibinfo {author} {\bibfnamefont {Y.}~\bibnamefont
  {Cui}}, \bibinfo {author} {\bibfnamefont {Y.}~\bibnamefont {Hikita}}, \ and\
  \bibinfo {author} {\bibfnamefont {H.~Y.}\ \bibnamefont {Hwang}},\ }\href
  {\doibase 10.1038/s41586-019-1496-5} {\bibfield  {journal} {\bibinfo
  {journal} {Nature}\ }\textbf {\bibinfo {volume} {572}},\ \bibinfo {pages}
  {624} (\bibinfo {year} {2019}{\natexlab{a}})}\BibitemShut {NoStop}%
\bibitem [{\citenamefont {Brahlek}\ \emph {et~al.}(2019)\citenamefont
  {Brahlek}, \citenamefont {Rimal}, \citenamefont {Ok}, \citenamefont
  {Mukherjee}, \citenamefont {Mazza}, \citenamefont {Lu}, \citenamefont {Lee},
  \citenamefont {Ward}, \citenamefont {Unocic}, \citenamefont {Eres},\ and\
  \citenamefont {Oh}}]{Brahlek2019}%
  \BibitemOpen
  \bibfield  {author} {\bibinfo {author} {\bibfnamefont {M.}~\bibnamefont
  {Brahlek}}, \bibinfo {author} {\bibfnamefont {G.}~\bibnamefont {Rimal}},
  \bibinfo {author} {\bibfnamefont {J.~M.}\ \bibnamefont {Ok}}, \bibinfo
  {author} {\bibfnamefont {D.}~\bibnamefont {Mukherjee}}, \bibinfo {author}
  {\bibfnamefont {A.~R.}\ \bibnamefont {Mazza}}, \bibinfo {author}
  {\bibfnamefont {Q.}~\bibnamefont {Lu}}, \bibinfo {author} {\bibfnamefont
  {H.~N.}\ \bibnamefont {Lee}}, \bibinfo {author} {\bibfnamefont {T.~Z.}\
  \bibnamefont {Ward}}, \bibinfo {author} {\bibfnamefont {R.~R.}\ \bibnamefont
  {Unocic}}, \bibinfo {author} {\bibfnamefont {G.}~\bibnamefont {Eres}}, \ and\
  \bibinfo {author} {\bibfnamefont {S.}~\bibnamefont {Oh}},\ }\href {\doibase
  10.1103/PhysRevMaterials.3.093401} {\bibfield  {journal} {\bibinfo  {journal}
  {Physical Review Materials}\ }\textbf {\bibinfo {volume} {3}},\ \bibinfo
  {pages} {093401} (\bibinfo {year} {2019})}\BibitemShut {NoStop}%
\bibitem [{\citenamefont {Shaw}\ \emph {et~al.}(2009)\citenamefont {Shaw},
  \citenamefont {Nembach}, \citenamefont {Silva}, \citenamefont {Russek},
  \citenamefont {Geiss}, \citenamefont {Jones}, \citenamefont {Clark},
  \citenamefont {Leo},\ and\ \citenamefont {Smith}}]{Shaw2009}%
  \BibitemOpen
  \bibfield  {author} {\bibinfo {author} {\bibfnamefont {J.~M.}\ \bibnamefont
  {Shaw}}, \bibinfo {author} {\bibfnamefont {H.~T.}\ \bibnamefont {Nembach}},
  \bibinfo {author} {\bibfnamefont {T.~J.}\ \bibnamefont {Silva}}, \bibinfo
  {author} {\bibfnamefont {S.~E.}\ \bibnamefont {Russek}}, \bibinfo {author}
  {\bibfnamefont {R.}~\bibnamefont {Geiss}}, \bibinfo {author} {\bibfnamefont
  {C.}~\bibnamefont {Jones}}, \bibinfo {author} {\bibfnamefont
  {N.}~\bibnamefont {Clark}}, \bibinfo {author} {\bibfnamefont
  {T.}~\bibnamefont {Leo}}, \ and\ \bibinfo {author} {\bibfnamefont {D.~J.}\
  \bibnamefont {Smith}},\ }\href {\doibase 10.1103/PhysRevB.80.184419}
  {\bibfield  {journal} {\bibinfo  {journal} {Physical Review B}\ }\textbf
  {\bibinfo {volume} {80}},\ \bibinfo {pages} {184419} (\bibinfo {year}
  {2009})}\BibitemShut {NoStop}%
\bibitem [{\citenamefont {Weller}\ \emph {et~al.}(1993)\citenamefont {Weller},
  \citenamefont {Br{\"{a}}ndle},\ and\ \citenamefont {Chappert}}]{Weller1993}%
  \BibitemOpen
  \bibfield  {author} {\bibinfo {author} {\bibfnamefont {D.}~\bibnamefont
  {Weller}}, \bibinfo {author} {\bibfnamefont {H.}~\bibnamefont
  {Br{\"{a}}ndle}}, \ and\ \bibinfo {author} {\bibfnamefont {C.}~\bibnamefont
  {Chappert}},\ }\href {\doibase 10.1016/0304-8853(93)91246-4} {\bibfield
  {journal} {\bibinfo  {journal} {Journal of Magnetism and Magnetic Materials}\
  }\textbf {\bibinfo {volume} {121}},\ \bibinfo {pages} {461} (\bibinfo {year}
  {1993})}\BibitemShut {NoStop}%
\bibitem [{\citenamefont {Li}\ \emph {et~al.}(2019{\natexlab{b}})\citenamefont
  {Li}, \citenamefont {Khim}, \citenamefont {Chang}, \citenamefont {Fu},
  \citenamefont {Nandi}, \citenamefont {Li}, \citenamefont {Yang},
  \citenamefont {Blake}, \citenamefont {Parkin}, \citenamefont {Auffermann},
  \citenamefont {Sun}, \citenamefont {Muller}, \citenamefont {Mackenzie},\ and\
  \citenamefont {Felser}}]{Li2019Felser}%
  \BibitemOpen
  \bibfield  {author} {\bibinfo {author} {\bibfnamefont {G.}~\bibnamefont
  {Li}}, \bibinfo {author} {\bibfnamefont {S.}~\bibnamefont {Khim}}, \bibinfo
  {author} {\bibfnamefont {C.~S.}\ \bibnamefont {Chang}}, \bibinfo {author}
  {\bibfnamefont {C.}~\bibnamefont {Fu}}, \bibinfo {author} {\bibfnamefont
  {N.}~\bibnamefont {Nandi}}, \bibinfo {author} {\bibfnamefont
  {F.}~\bibnamefont {Li}}, \bibinfo {author} {\bibfnamefont {Q.}~\bibnamefont
  {Yang}}, \bibinfo {author} {\bibfnamefont {G.~R.}\ \bibnamefont {Blake}},
  \bibinfo {author} {\bibfnamefont {S.}~\bibnamefont {Parkin}}, \bibinfo
  {author} {\bibfnamefont {G.}~\bibnamefont {Auffermann}}, \bibinfo {author}
  {\bibfnamefont {Y.}~\bibnamefont {Sun}}, \bibinfo {author} {\bibfnamefont
  {D.~A.}\ \bibnamefont {Muller}}, \bibinfo {author} {\bibfnamefont {A.~P.}\
  \bibnamefont {Mackenzie}}, \ and\ \bibinfo {author} {\bibfnamefont
  {C.}~\bibnamefont {Felser}},\ }\href {\doibase 10.1021/acsenergylett.9b01527}
  {\bibfield  {journal} {\bibinfo  {journal} {ACS Energy Letters}\ }\textbf
  {\bibinfo {volume} {4}},\ \bibinfo {pages} {2185} (\bibinfo {year}
  {2019}{\natexlab{b}})}\BibitemShut {NoStop}%
\bibitem [{\citenamefont {Kim}\ \emph {et~al.}(1997)\citenamefont {Kim},
  \citenamefont {Chernov}, \citenamefont {Kortright},\ and\ \citenamefont
  {Koo}}]{Kim1997}%
  \BibitemOpen
  \bibfield  {author} {\bibinfo {author} {\bibfnamefont {S.~K.}\ \bibnamefont
  {Kim}}, \bibinfo {author} {\bibfnamefont {V.~A.}\ \bibnamefont {Chernov}},
  \bibinfo {author} {\bibfnamefont {J.~B.}\ \bibnamefont {Kortright}}, \ and\
  \bibinfo {author} {\bibfnamefont {Y.~M.}\ \bibnamefont {Koo}},\ }\href
  {\doibase 10.1063/1.119470} {\bibfield  {journal} {\bibinfo  {journal}
  {Applied Physics Letters}\ }\textbf {\bibinfo {volume} {71}},\ \bibinfo
  {pages} {66} (\bibinfo {year} {1997})}\BibitemShut {NoStop}%
\bibitem [{\citenamefont {Harumoto}\ \emph {et~al.}(2019)\citenamefont
  {Harumoto}, \citenamefont {Shi},\ and\ \citenamefont
  {Nakamura}}]{Harumoto2019}%
  \BibitemOpen
  \bibfield  {author} {\bibinfo {author} {\bibfnamefont {T.}~\bibnamefont
  {Harumoto}}, \bibinfo {author} {\bibfnamefont {J.}~\bibnamefont {Shi}}, \
  and\ \bibinfo {author} {\bibfnamefont {Y.}~\bibnamefont {Nakamura}},\ }\href
  {\doibase 10.1063/1.5111649} {\bibfield  {journal} {\bibinfo  {journal}
  {Journal of Applied Physics}\ }\textbf {\bibinfo {volume} {126}},\ \bibinfo
  {pages} {083906} (\bibinfo {year} {2019})}\BibitemShut {NoStop}%
\bibitem [{\citenamefont {Takahashi}\ \emph {et~al.}(1993)\citenamefont
  {Takahashi}, \citenamefont {Tsunashima}, \citenamefont {Iwata},\ and\
  \citenamefont {Uchiyama}}]{Takahashi1993}%
  \BibitemOpen
  \bibfield  {author} {\bibinfo {author} {\bibfnamefont {H.}~\bibnamefont
  {Takahashi}}, \bibinfo {author} {\bibfnamefont {S.}~\bibnamefont
  {Tsunashima}}, \bibinfo {author} {\bibfnamefont {S.}~\bibnamefont {Iwata}}, \
  and\ \bibinfo {author} {\bibfnamefont {S.}~\bibnamefont {Uchiyama}},\ }\href
  {\doibase 10.1016/0304-8853(93)90602-X} {\bibfield  {journal} {\bibinfo
  {journal} {Journal of Magnetism and Magnetic Materials}\ }\textbf {\bibinfo
  {volume} {126}},\ \bibinfo {pages} {282} (\bibinfo {year}
  {1993})}\BibitemShut {NoStop}%
\bibitem [{\citenamefont {Wu}\ \emph {et~al.}(1992)\citenamefont {Wu},
  \citenamefont {St{\"{o}}hr}, \citenamefont {Hermsmeier}, \citenamefont
  {Samant},\ and\ \citenamefont {Weller}}]{Wu1992}%
  \BibitemOpen
  \bibfield  {author} {\bibinfo {author} {\bibfnamefont {Y.}~\bibnamefont
  {Wu}}, \bibinfo {author} {\bibfnamefont {J.}~\bibnamefont {St{\"{o}}hr}},
  \bibinfo {author} {\bibfnamefont {B.~D.}\ \bibnamefont {Hermsmeier}},
  \bibinfo {author} {\bibfnamefont {M.~G.}\ \bibnamefont {Samant}}, \ and\
  \bibinfo {author} {\bibfnamefont {D.}~\bibnamefont {Weller}},\ }\href
  {\doibase 10.1103/PhysRevLett.69.2307} {\bibfield  {journal} {\bibinfo
  {journal} {Physical Review Letters}\ }\textbf {\bibinfo {volume} {69}},\
  \bibinfo {pages} {2307} (\bibinfo {year} {1992})}\BibitemShut {NoStop}%
\bibitem [{\citenamefont {Weller}\ \emph {et~al.}(1994)\citenamefont {Weller},
  \citenamefont {Wu}, \citenamefont {St{\"{o}}hr}, \citenamefont {Samant},
  \citenamefont {Hermsmeier},\ and\ \citenamefont {Chappert}}]{Weller1994}%
  \BibitemOpen
  \bibfield  {author} {\bibinfo {author} {\bibfnamefont {D.}~\bibnamefont
  {Weller}}, \bibinfo {author} {\bibfnamefont {Y.}~\bibnamefont {Wu}}, \bibinfo
  {author} {\bibfnamefont {J.}~\bibnamefont {St{\"{o}}hr}}, \bibinfo {author}
  {\bibfnamefont {M.~G.}\ \bibnamefont {Samant}}, \bibinfo {author}
  {\bibfnamefont {B.~D.}\ \bibnamefont {Hermsmeier}}, \ and\ \bibinfo {author}
  {\bibfnamefont {C.}~\bibnamefont {Chappert}},\ }\href {\doibase
  10.1103/PhysRevB.49.12888} {\bibfield  {journal} {\bibinfo  {journal}
  {Physical Review B}\ }\textbf {\bibinfo {volume} {49}},\ \bibinfo {pages}
  {12888} (\bibinfo {year} {1994})}\BibitemShut {NoStop}%
\bibitem [{\citenamefont {Hong}\ \emph {et~al.}(2005)\citenamefont {Hong},
  \citenamefont {Sankar}, \citenamefont {Berkowitz},\ and\ \citenamefont
  {Egelhoff}}]{Hong2005}%
  \BibitemOpen
  \bibfield  {author} {\bibinfo {author} {\bibfnamefont {J.~I.}\ \bibnamefont
  {Hong}}, \bibinfo {author} {\bibfnamefont {S.}~\bibnamefont {Sankar}},
  \bibinfo {author} {\bibfnamefont {A.~E.}\ \bibnamefont {Berkowitz}}, \ and\
  \bibinfo {author} {\bibfnamefont {W.~F.}\ \bibnamefont {Egelhoff}},\ }\href
  {\doibase 10.1016/j.jmmm.2004.07.054} {\bibfield  {journal} {\bibinfo
  {journal} {Journal of Magnetism and Magnetic Materials}\ }\textbf {\bibinfo
  {volume} {285}},\ \bibinfo {pages} {359} (\bibinfo {year}
  {2005})}\BibitemShut {NoStop}%
\bibitem [{\citenamefont {Kim}\ \emph {et~al.}(1993)\citenamefont {Kim},
  \citenamefont {Lee},\ and\ \citenamefont {Chung}}]{Kim1993}%
  \BibitemOpen
  \bibfield  {author} {\bibinfo {author} {\bibfnamefont {S.}~\bibnamefont
  {Kim}}, \bibinfo {author} {\bibfnamefont {S.~R.}\ \bibnamefont {Lee}}, \ and\
  \bibinfo {author} {\bibfnamefont {J.~D.}\ \bibnamefont {Chung}},\ }\href
  {\doibase 10.1063/1.352643} {\bibfield  {journal} {\bibinfo  {journal}
  {Journal of Applied Physics}\ }\textbf {\bibinfo {volume} {73}},\ \bibinfo
  {pages} {6344} (\bibinfo {year} {1993})}\BibitemShut {NoStop}%
\bibitem [{\citenamefont {Aoki}\ \emph {et~al.}(1996)\citenamefont {Aoki},
  \citenamefont {Honda}, \citenamefont {Sato}, \citenamefont {Kobayashi},
  \citenamefont {Hashimoto}, \citenamefont {Yokoyama},\ and\ \citenamefont
  {Hanyu}}]{Aoki1996}%
  \BibitemOpen
  \bibfield  {author} {\bibinfo {author} {\bibfnamefont {Y.}~\bibnamefont
  {Aoki}}, \bibinfo {author} {\bibfnamefont {K.}~\bibnamefont {Honda}},
  \bibinfo {author} {\bibfnamefont {H.}~\bibnamefont {Sato}}, \bibinfo {author}
  {\bibfnamefont {Y.}~\bibnamefont {Kobayashi}}, \bibinfo {author}
  {\bibfnamefont {S.}~\bibnamefont {Hashimoto}}, \bibinfo {author}
  {\bibfnamefont {T.}~\bibnamefont {Yokoyama}}, \ and\ \bibinfo {author}
  {\bibfnamefont {T.}~\bibnamefont {Hanyu}},\ }\href {\doibase
  10.1016/0304-8853(96)00077-7} {\bibfield  {journal} {\bibinfo  {journal}
  {Journal of Magnetism and Magnetic Materials}\ }\textbf {\bibinfo {volume}
  {162}},\ \bibinfo {pages} {1} (\bibinfo {year} {1996})}\BibitemShut {NoStop}%
\bibitem [{\citenamefont {Rosenblatt}\ \emph {et~al.}(2010)\citenamefont
  {Rosenblatt}, \citenamefont {Karpovski},\ and\ \citenamefont
  {Gerber}}]{Rosenblatt2010}%
  \BibitemOpen
  \bibfield  {author} {\bibinfo {author} {\bibfnamefont {D.}~\bibnamefont
  {Rosenblatt}}, \bibinfo {author} {\bibfnamefont {M.}~\bibnamefont
  {Karpovski}}, \ and\ \bibinfo {author} {\bibfnamefont {A.}~\bibnamefont
  {Gerber}},\ }\href {\doibase 10.1063/1.3291707} {\bibfield  {journal}
  {\bibinfo  {journal} {Applied Physics Letters}\ }\textbf {\bibinfo {volume}
  {96}},\ \bibinfo {pages} {022512} (\bibinfo {year} {2010})}\BibitemShut
  {NoStop}%
\bibitem [{\citenamefont {Keskin}\ \emph {et~al.}(2013)\citenamefont {Keskin},
  \citenamefont {Aktaş}, \citenamefont {Schmalhorst}, \citenamefont {Reiss},
  \citenamefont {Zhang}, \citenamefont {Weischenberg},\ and\ \citenamefont
  {Mokrousov}}]{Keskin2013}%
  \BibitemOpen
  \bibfield  {author} {\bibinfo {author} {\bibfnamefont {V.}~\bibnamefont
  {Keskin}}, \bibinfo {author} {\bibfnamefont {B.}~\bibnamefont {Aktaş}},
  \bibinfo {author} {\bibfnamefont {J.}~\bibnamefont {Schmalhorst}}, \bibinfo
  {author} {\bibfnamefont {G.}~\bibnamefont {Reiss}}, \bibinfo {author}
  {\bibfnamefont {H.}~\bibnamefont {Zhang}}, \bibinfo {author} {\bibfnamefont
  {J.}~\bibnamefont {Weischenberg}}, \ and\ \bibinfo {author} {\bibfnamefont
  {Y.}~\bibnamefont {Mokrousov}},\ }\href {\doibase 10.1063/1.4776737}
  {\bibfield  {journal} {\bibinfo  {journal} {Applied Physics Letters}\
  }\textbf {\bibinfo {volume} {102}},\ \bibinfo {pages} {022416} (\bibinfo
  {year} {2013})}\BibitemShut {NoStop}%
\bibitem [{\citenamefont {Winer}\ \emph {et~al.}(2015)\citenamefont {Winer},
  \citenamefont {Segal}, \citenamefont {Karpovski}, \citenamefont {Shelukhin},\
  and\ \citenamefont {Gerber}}]{Winer2015}%
  \BibitemOpen
  \bibfield  {author} {\bibinfo {author} {\bibfnamefont {G.}~\bibnamefont
  {Winer}}, \bibinfo {author} {\bibfnamefont {A.}~\bibnamefont {Segal}},
  \bibinfo {author} {\bibfnamefont {M.}~\bibnamefont {Karpovski}}, \bibinfo
  {author} {\bibfnamefont {V.}~\bibnamefont {Shelukhin}}, \ and\ \bibinfo
  {author} {\bibfnamefont {A.}~\bibnamefont {Gerber}},\ }\href {\doibase
  10.1063/1.4935023} {\bibfield  {journal} {\bibinfo  {journal} {Journal of
  Applied Physics}\ }\textbf {\bibinfo {volume} {118}},\ \bibinfo {pages}
  {173901} (\bibinfo {year} {2015})}\BibitemShut {NoStop}%
\bibitem [{\citenamefont {Nagai}\ \emph {et~al.}(2005)\citenamefont {Nagai},
  \citenamefont {Shirakawa}, \citenamefont {Ikeda}, \citenamefont {Iwasaki},
  \citenamefont {Nishimura},\ and\ \citenamefont {Kosaka}}]{Nagai2005}%
  \BibitemOpen
  \bibfield  {author} {\bibinfo {author} {\bibfnamefont {I.}~\bibnamefont
  {Nagai}}, \bibinfo {author} {\bibfnamefont {N.}~\bibnamefont {Shirakawa}},
  \bibinfo {author} {\bibfnamefont {S.~I.}\ \bibnamefont {Ikeda}}, \bibinfo
  {author} {\bibfnamefont {R.}~\bibnamefont {Iwasaki}}, \bibinfo {author}
  {\bibfnamefont {H.}~\bibnamefont {Nishimura}}, \ and\ \bibinfo {author}
  {\bibfnamefont {M.}~\bibnamefont {Kosaka}},\ }\href {\doibase
  10.1063/1.1992671} {\bibfield  {journal} {\bibinfo  {journal} {Applied
  Physics Letters}\ }\textbf {\bibinfo {volume} {87}},\ \bibinfo {pages}
  {024105} (\bibinfo {year} {2005})}\BibitemShut {NoStop}%
\bibitem [{\citenamefont {K{\"{a}}stle}\ \emph {et~al.}(2004)\citenamefont
  {K{\"{a}}stle}, \citenamefont {Boyen}, \citenamefont {Schr{\"{o}}der},
  \citenamefont {Plettl},\ and\ \citenamefont {Ziemann}}]{Kastle2004}%
  \BibitemOpen
  \bibfield  {author} {\bibinfo {author} {\bibfnamefont {G.}~\bibnamefont
  {K{\"{a}}stle}}, \bibinfo {author} {\bibfnamefont {H.~G.}\ \bibnamefont
  {Boyen}}, \bibinfo {author} {\bibfnamefont {A.}~\bibnamefont
  {Schr{\"{o}}der}}, \bibinfo {author} {\bibfnamefont {A.}~\bibnamefont
  {Plettl}}, \ and\ \bibinfo {author} {\bibfnamefont {P.}~\bibnamefont
  {Ziemann}},\ }\href {\doibase 10.1103/PhysRevB.70.165414} {\bibfield
  {journal} {\bibinfo  {journal} {Physical Review B}\ }\textbf {\bibinfo
  {volume} {70}},\ \bibinfo {pages} {165414} (\bibinfo {year}
  {2004})}\BibitemShut {NoStop}%
\bibitem [{\citenamefont {Chawla}\ and\ \citenamefont
  {Gall}(2012)}]{Chawla2012}%
  \BibitemOpen
  \bibfield  {author} {\bibinfo {author} {\bibfnamefont {J.~S.}\ \bibnamefont
  {Chawla}}\ and\ \bibinfo {author} {\bibfnamefont {D.}~\bibnamefont {Gall}},\
  }\href {\doibase 10.1063/1.3684976} {\bibfield  {journal} {\bibinfo
  {journal} {Journal of Applied Physics}\ }\textbf {\bibinfo {volume} {111}},\
  \bibinfo {pages} {043708} (\bibinfo {year} {2012})}\BibitemShut {NoStop}%
\bibitem [{\citenamefont {{de Vries}}(1988)}]{deVries1988}%
  \BibitemOpen
  \bibfield  {author} {\bibinfo {author} {\bibfnamefont {J.~W.}\ \bibnamefont
  {{de Vries}}},\ }\href {\doibase 10.1016/0040-6090(88)90478-6} {\bibfield
  {journal} {\bibinfo  {journal} {Thin Solid Films}\ }\textbf {\bibinfo
  {volume} {167}},\ \bibinfo {pages} {25} (\bibinfo {year} {1988})}\BibitemShut
  {NoStop}%
\end{thebibliography}

%


\pagebreak

\end{document}